# Energy cycle for the Lorenz attractor

Vinicio Pelino,
Filippo Maimone

*Italian Air Force, CNMCA*
*Aeroporto "De Bernardi", Via di Pratica Di Mare,*
*I-00040 Pratica di Mare (Roma) Italy*

In this note we study energetics of Lorenz-63 system through its Lie-Poisson structure.

## I. INTRODUCTION

In 1955 E. Lorenz [1] introduced the concept of energy cycle as a powerful instrument to understand the nature of atmospheric circulation. In that context conversions between potential, kinetic and internal energy of a fluid were studied using atmospheric equations of motion under the action of an external radiative forcing and internal dissipative processes. Following these ideas, in this paper we will illustrate that chaotic dynamics governing Lorenz-63 model can be described introducing an appropriate energy cycle whose components are kinetic, potential energy and Casimir function derived from Lie-Poisson structure hidden in the system; Casimir functions, like enstrophy or potential vorticity in fluid dynamical context, are very useful in analysing stability conditions and global description of a dynamical system.
A typical equation describing dissipative-forced dynamical systems can be written in Einstein notation as:

$$\dot{x}_i = \{x_i, H\} - \Lambda_{ij} x_j + f_i \qquad i = 1,2...n \qquad (1)$$

Equations (1) have been written by Kolmogorov, as reported in [2], in a fluid dynamical context, but they are very common in simulating natural processes as useful in chaos synchronization [3]. Here, antisymmetric brackets represent the algebraic structure of Hamiltonian part of a system described by function $H$, and a cosymplectic matrix $\mathbf{J}$ [4],

$$\{F,G\} = J_{ik} \partial_i F \partial_k G . \qquad (2)$$

Positive definite diagonal matrix $\Lambda$ represents dissipation and the last term $\mathbf{f}$ represents external forcing. Such a formalism, as mentioned before, is particularly useful in fluid dynamics [5], where Navier-Stokes equations show interesting properties in their Hamiltonian part (Euler equations). Moreover, finite dimensional systems as (1) represent the proper reduction of fluid dynamical equations [6], in terms of conservation of the symplectic structures in the infinite domain [7]. Method of reduction, contrary to the classical truncation one, leads to the study of dynamics on Lie algebras, i.e to the study of Lie-Poisson equations on them, which are extremely interesting from the physical viewpoint and with a mathematical aesthetical appeal [8,9]. Given a group $\mathbf{G}$ and a real-valued function (possibly time dependent), $H: T_e^* \mathbf{G} \to \mathbf{R}$, which plays the role of the Hamiltonian, in the local co-ordinates $x_i$ the Lie-Poisson equations read as

$$\dot{x}_i = C_{ik}^j x_j \partial_k H, \quad (3)$$

where tensor $C_{ik}^j$ represents the constants of structure of the Lie algebra **g** and the cosymplectic matrix assumes the form $J_{ik}(\mathbf{x}) = C_{ik}^j x_j$. It is straightforward to show that, in this formalism **g** is endowed with a Poisson bracket characterized by expression (2) for functions $F, G \in C^{\infty}(\mathbf{g}^*)$. Casimir functions $C$ are given by the kernel of bracket (2), i.e. $\{C, G\} = 0, \forall G \in C^{\infty}(\mathbf{g}^*)$, therefore they represent the constants of motion of the Hamiltonian system, $\dot{C} = \{C, H\} = 0$; moreover they define a foliation of the phase space [10].

## II. LORENZ SYSTEM AND ITS GEOMETRY

Here we will be interested in $G = SO(3)$ with $J_{ij} = \varepsilon^i_{jk} x_i$, where $\varepsilon_{ijk}$ stands for the Levi-Civita symbol; in case of a quadratic Hamiltonian function,

$$K = \frac{1}{2}\Omega_{ik} x_i x_k \quad (4)$$

they represent the Euler equations for the rigid body, with Casimir $C = \frac{1}{2}\delta_{ij} x_i x_j$ and relative foliation geometry $\mathbf{S}^2$.

In a previous paper [11] it has been shown that also the famous Lorenz-63 system [12]

$$\begin{cases} \dot{x}_1 = -\sigma x_1 + \sigma x_2 \\ \dot{x}_2 = -x_1 x_3 + \rho x_1 - x_2 \\ \dot{x}_3 = x_1 x_2 - \beta x_3 \end{cases} \quad (5)$$

where $\sigma = 10, \beta = 8/3, \rho = 28$, can be written in the Kolmogorov formalism as in (1),

$$\begin{cases} \dot{x}_1 = -\sigma x_1 + \sigma x_2 \\ \dot{x}_2 = -x_1 x_3 - \sigma x_1 - x_2 \\ \dot{x}_3 = x_1 x_2 - \beta x_3 - \beta(\rho + \sigma) \end{cases} \quad (6)$$

Assuming the following axially symmetric gyrostat [13] Hamiltonian with rotational kinetic energy $K$ and a linear potential $U(x_k) = \omega_k x_k$ will be written as

$$H = K + U \quad (7)$$

with: $\Omega = \begin{bmatrix} \Omega_1 = 1 & 0 & 0 \\ 0 & \Omega_2 = 2 & 0 \\ 0 & 0 & \Omega_3 = 2 \end{bmatrix}$ inertia tensor, $\Lambda = \begin{bmatrix} \Lambda_1 = \sigma & 0 & 0 \\ 0 & \Lambda_2 = 1 & 0 \\ 0 & 0 & \Lambda_3 = \beta \end{bmatrix}$ dissipation, internal forcing given by an axisymmetric rotor $\boldsymbol{\omega} = [0, 0, \sigma]$ and external forcing

$\mathbf{f} = [0, 0, -\beta(\rho+\sigma)]$. In order to distinguish effects of different terms in the energy cycle we leave the notation $\boldsymbol{\omega} = [0, 0, \omega]$ assuming that for numerical values in Lorenz attractor $\omega = \sigma$. Casimir function $C(t)$, that is constant in the conservative case, will give a useful geometrical vision to the understanding of dynamical behaviour of (6). This is given by studying fixed points of $\dot{C}(t) = \dot{x}_i \partial_i C$, which defines an invariant triaxial ellipsoid $\Xi_0$ with center $\left\{0, 0, -\frac{\rho+\sigma}{2}\right\}$ and axes $a = f/2\sqrt{\beta\sigma}, b = f/2\sqrt{\beta}, c = f/2\beta$, having equation

$$-\Lambda_{ij} x_i x_j + f_i x_i = 0 \ . \tag{8}$$

Fixed points of system (6) $\mathbf{x}_\pm = \left\{\pm\sqrt{\beta(\rho-1)}, \pm\sqrt{\beta(\rho-1)}, -\sigma-1\right\}$ and $\mathbf{x}_0 = \{0, 0, -(\rho+\sigma)\}$ belong to $\Xi_0$. Computing the flux density of vector field $\mathbf{u} = \dot{\mathbf{x}}$ through this manifold $\varphi(\mathbf{u}) = u_i \partial_i \Xi$, because of reflection symmetry $x_i \to -x_i$, $i = 1, 2$ of equations (6), two symmetric regions are identified by respectively, $\varphi < 0$ and $\varphi > 0$. Lorenz attractor $\Psi$ intersects the manifold in these regions at maxima and minima of Casimir function, entering the ellipsoid trough $\min(C)$ twice where $\varphi < 0$, right $\Psi_R$ and left lobe $\Psi_L$, and symmetrically exiting trough $\max(C)$ twice where $\varphi > 0$, $\Psi \cap \Xi_0 = \{\min(C), \max(C)\}$ (**Fig.1**).

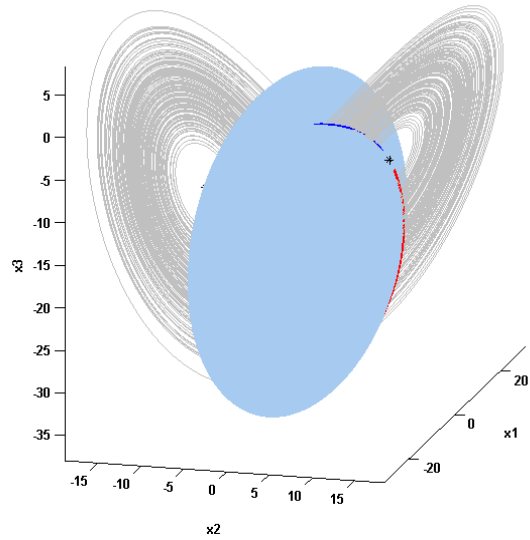

**Fig.1** Invariant ellipsoid $\Xi_0$ intersecting the attractor in the set $\Psi \cap \Xi_0 = \{\min(C), \max(C)\}$. Casimir maxima (red) and minima (blue) are shown. Black stars represent 2 of the 3 fixed points. Note that $\mathbf{x}_0$ lies on the southern pole. In order to show points on $\Xi_0$, only part of trajectory is shown.

The particular choice of parameters $\sigma = 10, \rho = 28, \beta = \frac{8}{3}$ has moreover the peculiarity for $\max(C)$ to be an ordered set [9]. This property gives the opportunity to find a range of variability for the *maximum* radius of the Casimir sphere $r(t) = \sqrt{C}$,

$$R_{\min} = \|R_\pm\| \leq R(t) \leq \|R_0\| = R_{\max}, \qquad (9)$$

where $\|R_\pm\| = \sqrt{2\beta(\rho-1) + (1+\sigma)^2}$ and $\|R_0\| = \rho + \sigma$; moreover ellipsoid $\Xi_0$ defines a natural Poincarè section for Lorenz equations with interesting properties in the associated return map for Casimir maxima [14].

## III. ENERGY CYCLE

In order to introduce an energy cycle into system (1), we first consider pure conservative case. Here dynamics lies on $\mathbf{S}^2$ and we note that introduction of potential $U$ produces a deformation on the geodesics trajectory [15] of the Riemannian metric given by the quadratic form (4). Adding the spin rotor $\omega x_3$, the centre of the ellipsoid of revolution is shifted from the origin, that remains the centre of the Casimir sphere of radius $\sqrt{2C}$. As regards fixed points, given at $\omega = 0$ by two isolated centers, namely the poles $(\pm\sqrt{2C},0,0)$ of $\mathbf{S}^2$, and all points belonging to 'equatorial circle' $(0, \sqrt{2C}\sin\theta, \sqrt{2C}\cos\theta)$, introduction of potential $U(x_3)$ reduces this set to four equilibrium points located at $F_{1,2} = (0,0,\pm\sqrt{2C})$ and $F_{3,4} = (\pm\sqrt{2C-\omega^2}, 0, -\omega)$. This last pair starts to migrate into the south pole of $\mathbf{S}^2$ as $\omega$ grows, and disappears for $\omega > \sqrt{2C}$ giving rise to an *oyster* bifurcation [16], leaving two stable centers $F_{1,2}$. Lie-Poisson structure of the system permits to analyze nonlinear stability of these two points introducing pseudoenergy functions [17] $I_{1,2} = H + \lambda_{1,2} C$, where $\lambda_{1,2}$ are the solutions of equation

$$\partial_i H \big|_{F_{1,2}} = -\lambda \partial_i C \big|_{F_{1,2}} \qquad (10)$$

at fixed points $F_{1,2} = (0,0,\pm\sqrt{2C})$. Computation of quadratic form $\partial_{i,j} I \big|_{F_{1,2}}$ shows $F_1$ as a maximum and $F_1$ as a minimum for $\omega > \sqrt{2C}$. We point out that introduction of potential reduces number of fixed points on the sphere to the minimal number 2; therefore it stabilizes the system's dynamics.

Total energy $E$, identified with Hamiltonian, does not change and in terms of $K$ and $U$, a simple energy cycle, similar to the classical oscillator, can be described using bracket formalism (4) with relative rules $\{U,H\} = \{U,K\} = \mathbf{C}(U,K)$ and $\{K,H\} = \{K,U\} = \mathbf{C}(K,U)$

$$\begin{cases} \dot{K} = \mathbf{C}(K,U) \\ \dot{U} = \mathbf{C}(U,K) \\ \dot{C} = 0 \end{cases} \qquad (11)$$

where $\mathbf{C}(U,K)$ is positive if energy is flowing from $U$ to $K$; the net rate of conversion of potential into kinetic energy factor is of the form

$$\mathbf{C}(U,K) = \omega(\Omega_1 - \Omega_2)x_1 x_2 \qquad (12)$$

due to the linear dependence of $U$ on $x_3$. As a result, a symmetry between quadrants I and III holds since $x_1 x_2 > 0 \Rightarrow \mathbf{C}(U,K) < 0$ corresponding to a net conversion of kinetic energy into potential one $K \to U$; the opposite happens in quadrants II and IV where $x_1 x_2 < 0 \Rightarrow \mathbf{C}(U,K) > 0$ and $U \to K$.

Following the ideas of extending the algebraic formalism of Hamiltonian dynamics to include dissipation [18], we introduce a Lyapunov function $L(x) \in C^\infty\left(so(3)^*\right)$, together with a symmetric bracket

$$\dot{F} = \langle L, F \rangle = g_{ik} \partial_i L \partial_k F \qquad (13)$$

where $g_{ik}(\mathbf{x})$ generally is a symmetric negative matrix. Taken alone, formalism (13) gives rise to a gradient system dynamics

$$\dot{x}_i = \langle L, x_i \rangle = g_{ik} \partial_k L \qquad (14).$$

Including Lie-Poisson structure (2), it is possible to study equations (11) adding various kinds of dissipation models depending on the choice of 'metric tensor' $g_{ik}(\mathbf{x})$

$$\begin{cases} \dot{K} = \mathbf{C}(K,U) + \langle L, K \rangle \\ \dot{U} = \mathbf{C}(U,K) + \langle L, U \rangle \\ \dot{C} = \langle L, C \rangle \end{cases} \qquad (15).$$

Because of $\mathbf{C}(C,H) = 0$ last equation in (15) describes the contraction of the manifold where motion takes place.

in order to find a dissipation process that naturally takes into account the compact and semisimple structure of $so(3)$, we use the so called Cartan-Killing dissipation derived form Cartan-Killing metric [18[

$$g_{ik} = \frac{1}{2} \varepsilon_{im}^n \varepsilon_{kn}^m \qquad (16)$$

Physically with this choice, for $L = \alpha C$ and $\alpha \in \mathbf{R}^+$, dynamics reduces to an isotropic linear damping $g_{ik} = -\delta_{ik}$ of both energy and Casimir functions, miming a Rayleigh dissipation; in this way trajectories approache a stable fixed point at the origin $\mathbf{x}_0 = (0,0,0)$.

Because of $\Lambda$ term of Lorenz equations, we can introduce a Lyapunov function with anisotropic dissipation of the form

$$L = \frac{1}{2}\Lambda_{ik} x_i x_k \qquad (17);$$

moreover in the spirit of formalism (13), an external torque representing forcing can be easily included in the symmetric bracket by a translation $L \to L + G$ where

$$G = -f \cdot x_3 \qquad (18).$$

It is interesting to note that on the ellipsoid $\Xi_0$, both $\nabla(L+G)$ and Lorenz field $\mathbf{u} = \dot{\mathbf{x}}$ are orthogonal to $\nabla C$, but since determinant $\|\nabla C, \nabla(L+G), \mathbf{u}\| \neq 0$, the 3- vectors do not belong to the same plane.

Energy cycle for Lorenz attractor can be finally written as

$$\begin{cases} \dot{K} = -\mathbf{C}(U,K) - \Lambda_{ij}\Omega_{jk} x_i x_k - \Omega_3 G \\ \dot{U} = \mathbf{C}(U,K) - \beta U + f\omega \\ \dot{C} = -(2L + G) \end{cases} \qquad (19),$$

i.e. defining forcing terms as

$$\begin{cases} F_K = \langle G, K \rangle = f\Omega_3 x_3 \\ F_U = \langle G, U \rangle = f\omega \\ F_C = \langle G, C \rangle = fx_3 \end{cases} \qquad (20).$$

In this formalism the first two equations of (19) describe energy variations of a particle dynamics constrained to move on a spherical surface of variable radius $r(t) = \sqrt{2C}$. It is easy to verify that for isotropic dissipation $L = \alpha C$, even in presence of forcing, equations (19) describe a purely dissipative dynamics. In spherical coordinates after simplification, it becomes

$$2\dot{r} = -2\alpha r - f\sin\theta \qquad (21).$$

For Lorenz parameters, combined effects of conservative part, anisotropic dissipation and forcing components of (1), makes dynamics of the spherical radius deterministic, bounded, recurrent and sensitive to initial conditions, as shown in **Fig.2**, in other words chaotic.

Motion on a variable but topologically stable manifold justifies the presence of last equation in (19) that takes into account the background field $C(t)$.

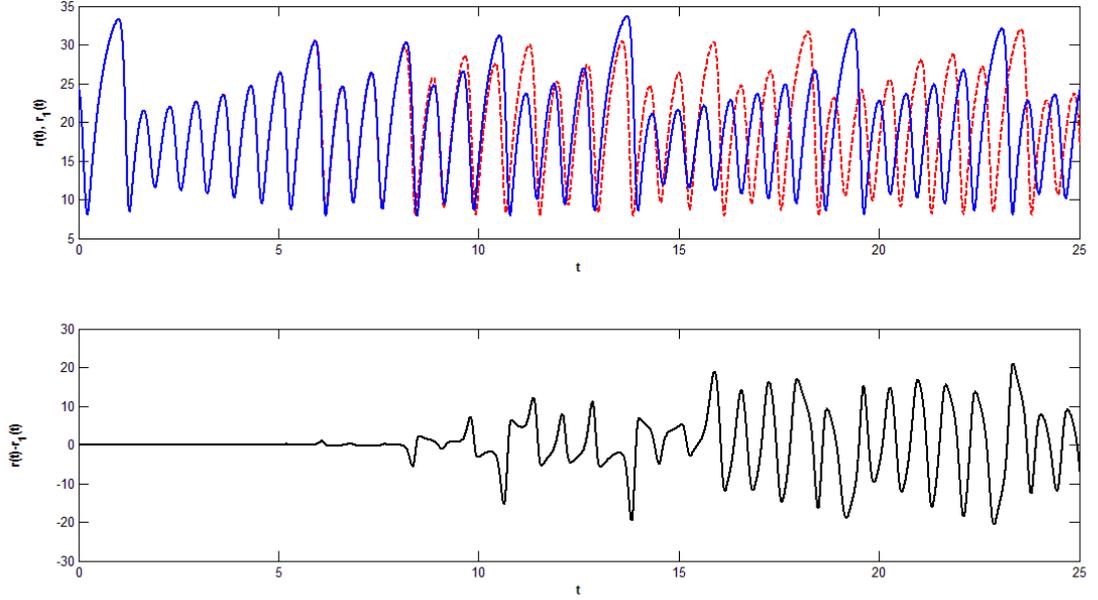

**Fig.2** Sensitivity to initial conditions for two numerically different Casimir functions Top: time evolution for two Casimir radii $r(t) = \sqrt{2C(t)}$. Bottom: $\Delta r(t)$; with $\Delta r(t_0) \approx 0.008$

If we consider steady state conditions, where all three time derivatives in (19) are set to zero, we note that last equation represents ellipsoid $\Xi_0$ of equation (8) and the only point solution lying on it is given by fixed point $\mathbf{x_0} = \{0, 0, -(\rho+\sigma)\}$ corresponding to the asymptotic $R_{max}$ in (9). Here $\mathbf{C}(U,K) = 0$, $K = 2C = (\rho+\sigma)^2$ and potential reaches its minimum value $U = -\sigma(\rho+\sigma)$.

In order to study behaviour of Casimir function and its conversion terms, it will be useful to rewrite last equation of (19) in terms of Lyapunov function $L$ and forcing $G$, both contained in a function $W(\mathbf{x}) = -(2L+G)$. As a matter of facts, substituting $\ddot{C} = -(2\dot{L}+\dot{G}) = \dot{W}$ in (19), we have:

$$\dot{W} = \mathbf{C}(W,K) + \mathbf{C}(W,U) + \langle L+G, W \rangle \qquad (22)$$

from which applying antisymmetric properties of Lie- Poisson bracket, conversion terms for $U$ and $K$ are written as

$$\mathbf{C}(W,K) = 2\mathbf{C}(K,L) + \mathbf{C}(K,G) = 2\left(\sum_i \varepsilon_{ijk} \Omega_j \Lambda_k\right) x_1 x_2 x_3 + f(\Omega_1 - \Omega_2) x_1 x_2$$

$$\mathbf{C}(W,U) = 2\mathbf{C}(U,L) + \mathbf{C}(U,G) = 2\omega(\sigma-1) x_1 x_2 + 0 \qquad (23)$$

Note that all of terms in (23) are linear functions of $C(U,K)$. For Lorenz parameters it results $\sum_i \varepsilon_{ijk}\Omega_j \Lambda_k > 0$, $\omega(\sigma-1) > 0$ and $f(\Omega_1 - \Omega_2) > 0$ from which energy cycle reads as

$$\begin{cases} \dot{K} = -C(U,K) - \Lambda_{ij}\Omega_{jk}x_i x_k - \Omega_3 G \\ \dot{U} = C(U,K) - \beta U + f\omega \\ \dot{W} = \frac{1}{\Omega_1 - \Omega_2}\left(2\frac{\left[(1-\beta)\Omega_1 + (\beta-\sigma)\Omega_2 + (\sigma-1)\Omega_3\right]}{\omega^2}U + 2(\sigma-1) + \frac{f}{\omega}(\Omega_1 - \Omega_2)\right) \cdot C(U,K) + 2\|L\|^2 + \|G\|^2 + 3LG \end{cases}$$

(24)

This reformulation of energy cycle takes into account dissipation and forcing in conversion terms between $K, U, W$ where $W = \dot{C}$ at least formally plays the role of internal energy of the system.

## IV. CONVERSION FACTORS

Apart from dissipation and forcing terms from (24) it is clear that energy cycle for Lorenz attractor can be studied by analysing the behaviour of conversion term $C(U,K)$ shown in **Fig.3**.

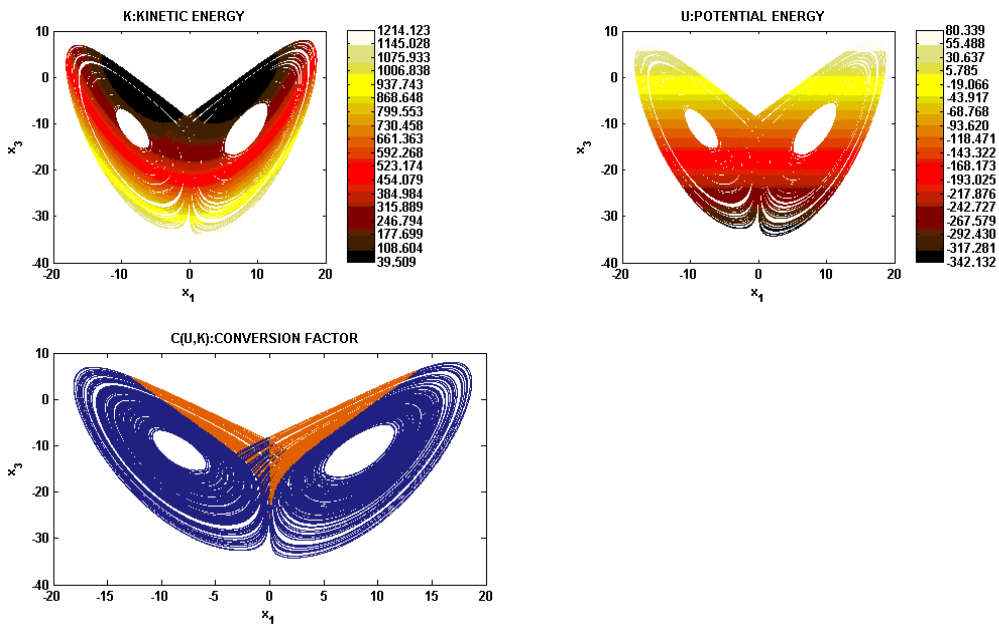

**Fig.3**. Kinetic energy, potential energy and their relative conversion term $C(U,K)$ are plotted on Lorenz attractor. Colour ranges from black to bright copper, going from low to high numerical values. For $C(U,K)$, regions of negative values $x_1 x_2 < 0$, or transition regions where $C(U,K) > 0$, are shown in orange. They cover about 12% of the attractor.

Given that dynamics of Lorenz system is well described as a sequence of traps and jumps between the two lobes of the strange attractor; we observe what follows.

When trapped in a lobe, system state experiences spiral-like trajectory, whose centre is in a fixed point $\mathbf{x}_\pm = \{\pm\sqrt{\beta(\rho-1)}, \pm\sqrt{\beta(\rho-1)}, -\sigma-1\}$ and whose radius increases in time as energy and Casimir maxima, till touching the boundaries planes $x_1 = 0$ or $x_2 = 0$ where a transition to the opposite lobe occurs.

After lobe transition, trajectory starts from regions close to the opposite unstable fixed point $\mathbf{x}_\mp = \{\mp\sqrt{\beta(\rho-1)}, \mp\sqrt{\beta(\rho-1)}, -\sigma-1\}$, moving towards boundary planes and so on.

Looking at the attractor as a fractal object, 12% of its whole structure lies in transition regions where $x_1 x_2 < 0$.

The above described dynamical behaviour above described is ruled by laws (23) and (24) as shown in **Fig.4**.

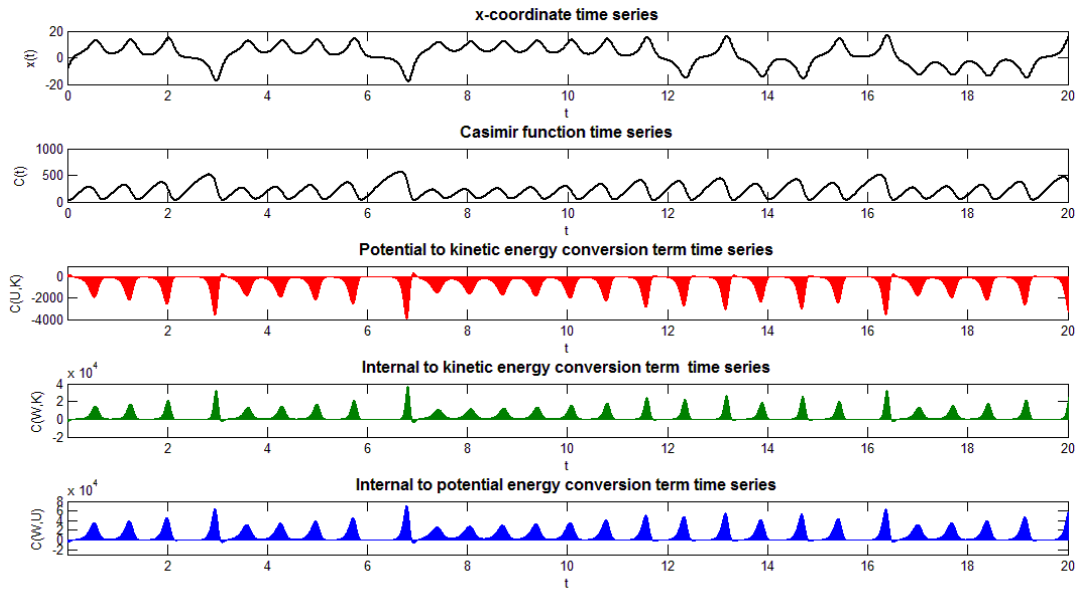

**Fig.4** Time evolution for energy cycle variables. Top sign of $x_1(t)$ represents jumps and traps. Jumps are associated to spikes in conversion terms. Second row: $C(t)$ expansions and contractions of the Casimir sphere, in its maxima, are modulated by the sign of conversion terms values. Middle and fourth lines, conversion terms $\mathbf{C}(U,K)$ and $\mathbf{C}(W,K)$ show an opposite phase dynamics. Bottom $\mathbf{C}(W,U)$. Note inequality $|\mathbf{C}(U,K)| < |\mathbf{C}(W,U)| < |\mathbf{C}(W,K)|$

Reminding that $\mathbf{C}(W,U) = 2\mathbf{C}(U,L)$, a conversion from kinetic to potential energy $K \to U$ occurs in trapping regions $x_1 x_2 > 0$ where a number of loops around $\mathbf{x}_\pm$ of increasing radius occurs; in the same regions a conversion $W \to U$ occurs, (**Fig.5**).

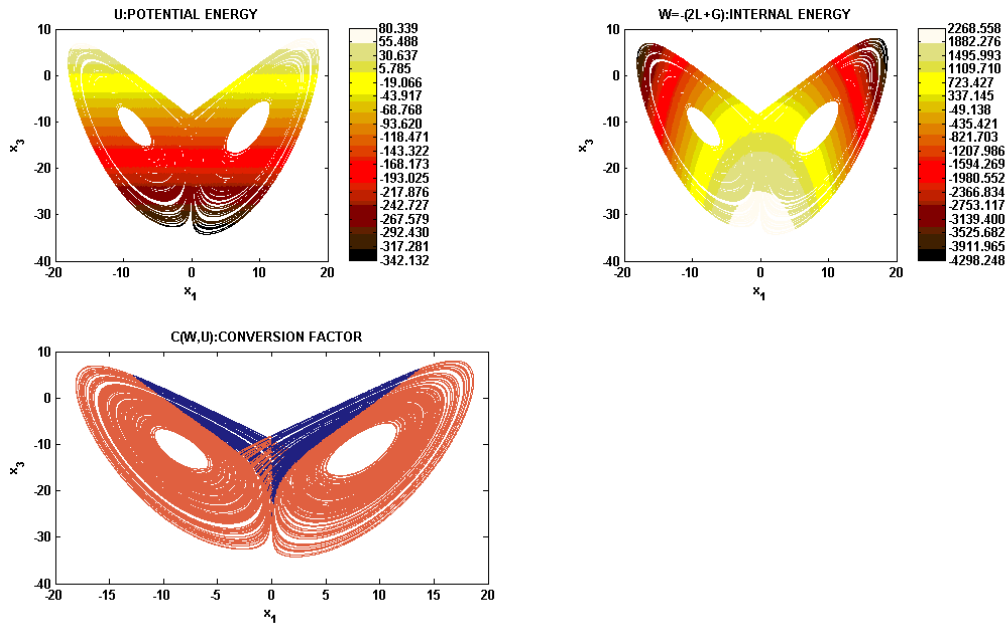

**Fig.5** Potential energy, "internal energy" and their relative conversion term $C(W,U)$ are plotted on Lorenz attractor. Colour ranges from black to bright copper, going from low to high numerical values. Regions where $C(W,U) > 0$, are shown in orange.

As regards to term $C(W,K)$, coordinate $x_3$ drives the behaviour of $C(K,L)$, giving rise to both conversions $K \leftrightarrow L$ inside a lobe of the attractor, while $K \to G$ is the only possible conversion, (**Fig.6**).

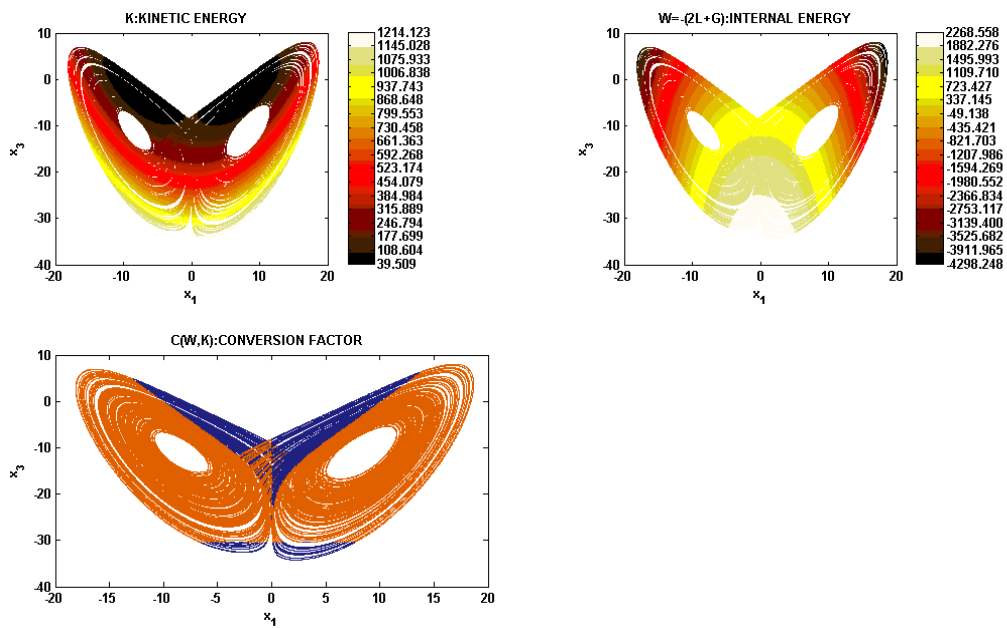

**Fig.6**. Kinetic energy, "internal energy" and their relative conversion term $C(W,K)$ are plotted on Lorenz attractor. Colour ranges from black to bright copper, going from low to high numerical values. Regions where $C(W,K) > 0$ are shown in orange.

In trapping regions function $W = \dot{C}$ act as a source for total energy $H = U + K$, $\mathbf{C}(W,K) > 0$ and $\mathbf{C}(W,U) > 0$.

In this phase, Casimir sphere, centred in $\mathbf{x_0} = \{0,0,0\}$, expands.

In regions $x_1 x_2 < 0$ a drastic change in energy cycle occurs, depending on the bounded nature of the system. Here potential energy is transformed into kinetic energy $U \to K$ and Lyapunov function into potential energy, $L \to U$; therefore an implosion of Casimir sphere occurs. also, because of (23) $L \leftrightarrow K$ and $G \to K$.

In lobe-jumping regions function $W = \dot{C}$ acts as a sink for both potential and kinetic energy, since $\mathbf{C}(W,K) < 0$ and $\mathbf{C}(W,U) < 0$.

Concerning numerical values of conversion terms, the following inequalities hold:

$$\left|\frac{\mathbf{C}(K,L)}{\mathbf{C}(K,G)}\right| = 2\left|\left(\sum_i \varepsilon_{ijk}\Omega_j \Lambda_k\right)\frac{x_3}{f}\right| < 1, \qquad (25)$$

$$\left|\frac{\mathbf{C}(K,U)}{\mathbf{C}(W,U)}\right| = \left|\frac{-1}{2(\sigma-1)}\right| < 1, \qquad (26)$$

$$\left|\frac{\mathbf{C}(W,K)}{\mathbf{C}(W,U)}\right| \leq \left|\frac{(3\beta-2)(\rho+\sigma)}{2\omega(\sigma-1)}\right| < 1, \qquad (27)$$

(since $\max|z| = \rho + \sigma$):

$$|\mathbf{C}(K,U)| < |\mathbf{C}(W,K)| < |\mathbf{C}(W,U)|. \qquad (28)$$

Conversion rules are reassumed in the following diagram where dashed lines represent cycle in jumping regions $x_1 x_2 < 0$ and continuous lines refer to trapping lobes $x_1 x_2 > 0$ of the attractor.

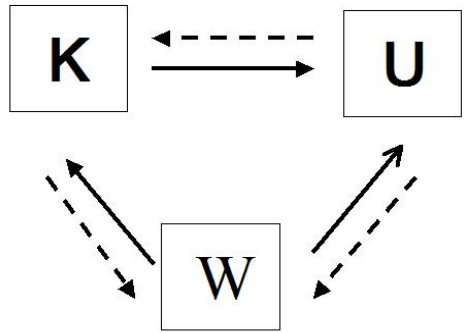

## V. MECHANICAL APE\UPE AND PREDICTABILITY

In the spirit of 1955 Lorenz work on general circulation of the atmosphere [1], we introduce for system (6) quantities respectively known as available potential energy $APE = U_{max} - U_{min}$ and unavailable potential energy $UPE = U_{min}$; they represent, respectively, the portion of potential energy that can be converted into kinetic energy, and the portion that cannot.

In atmospheric science $APE$ is a very important subject since its variability determines transitions in the atmospheric circulation. In analogy with the theory of Margules [19], in which a fluid inside a vessel is put into motion and free surface oscillates up and down, while potential energy is constrained from below by the potential energy of the fluid at rest we start by considering the conservative system :

$$\begin{cases} \dot{x}_1 = \sigma x_2 \\ \dot{x}_2 = -x_1 x_3 - \sigma x_1 \\ \dot{x}_3 = x_1 x_2 \end{cases} \qquad (29)$$

fixing Casimir and Energy values $(C_0, E_0)$, equation $\mathbf{C}(U, K) = 0$ gives at $x_1 = 0 \Rightarrow K_{max} = \Omega C_0$ and $U_{min} = E_0 - \Omega C$; at $x_2 = 0 \Rightarrow U_{max} = \left[ -\omega^2 + \omega\sqrt{\omega^2 - 2(\Omega - 1)(C_0 - E_0)} \right] / (\Omega - 1)$ and $K_{min} = E_0 - U_{max}$, where $\Omega = \Omega_2 = \Omega_3$ for Lorenz system, therefore we get:

$$\begin{cases} APE = \dfrac{-\omega^2 + \omega\sqrt{\omega^2 - 2(\Omega - 1)(C_0 - E_0)}}{\Omega - 1} - E_0 + \Omega C_0 \\ UPE = E_0 - \Omega C_0 \end{cases} \qquad (30)$$

In case of full Lorenz system, energy and Casimir are not conserved even though their associated surfaces intersect instantaneously; therefore introducing dissipation and forcing one can still consider the evolution of $APE$ and $UPE$ as state functions. **Fig .7** shows a graphical representation of the two quantities over Lorenz attractor; $APE$ increases as $x_3$ decreases while $UPE$ follows the opposite way.

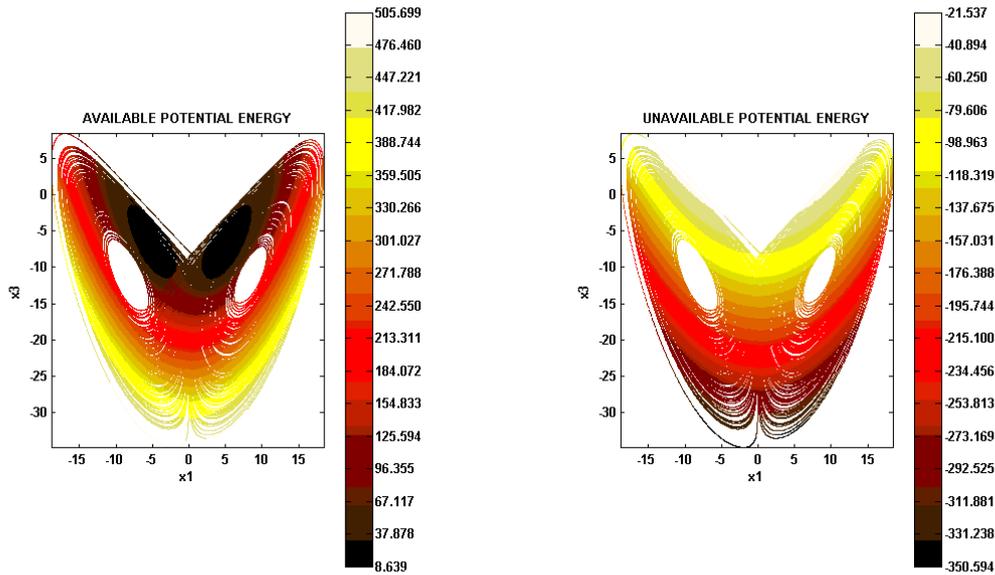

**Fig.7** Mechanical APE and UPE plotted on Lorenz attractor.

This behaviour coincides with that of predictability regions over the attractor computed using breeding vectors technique [20] and shown in **Fig.8**. Giving an initial perturbation $\delta \mathbf{x}_0$, red vector growth $g$ over n=8 steps is computed as $g = \frac{1}{n}\log\left(|\delta \mathbf{x}|/\delta \mathbf{x}_0\right)$, regions of the Lorenz attractor within which all infinitesimal uncertainties decrease with time [21] are located in regions of $\max(UPE)$ and $\min(APE)$.

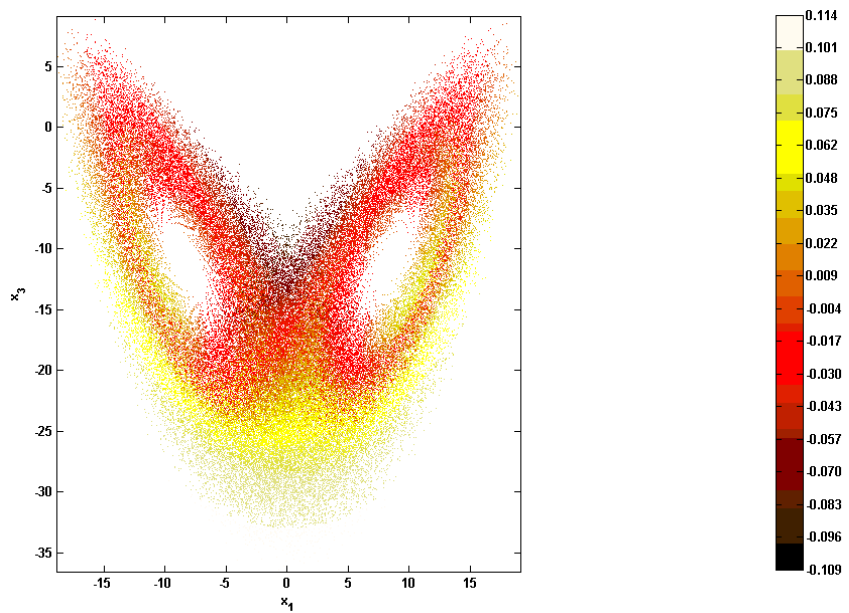

**Fig.8** Breeding vectors map for Lorenz attractor, solution of equations (6). Colour ranges from black to bright copper meaning low to high predictability.

Finally we note that the set $\Psi \cap \Xi_{APE} = \{\min(APE), \max(APE)\}$, where $\Xi_{APE}$ represents the surface $\frac{d}{dt}(APE) = 0$ has as ordered subset $\max(APE)$. In this view, Lorenz map for $APE$ maxima $APE$ shown in **Fig.9** assumes the meaning of 'handbook' for regime transitions; for low values of $APE$ system is trapped withine one lobe until it reaches the minimum necessary value to jump into the opposite one.

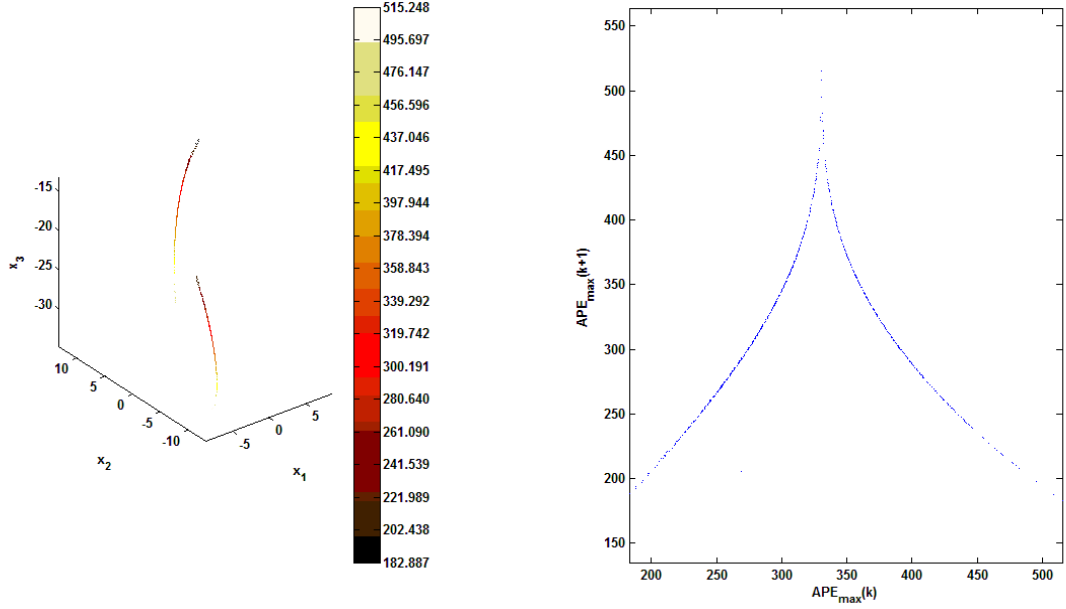

**Fig.9** Left: ordered set for APE maximum values lying on ellipsoid $\Xi_{APE}$ (not shown); colour ranges from black to bright copper, going from low to high numerical values. Right: Lorenz map for APE maxima.

## VI. ENERGETICS AND DYNAMICS

Physically, introduction of function $W$ has the following meaning: ellipsoid $\Xi_0$ contains sets $\min_{\Psi_{L,R}}(C)$ and $\max_{\Psi_{L,R}}(C)$ for system (6) and acts as a boundary for regions of maximum forcing and dissipation.

Inside solid ellipsoid $\Xi$, $G > 2L$ and forcing drives motion, also because of $W = \dot{C} > 0$ which implies that Casimir sphere continually expands. and for $C(U, K) < 0$ trajectory of (6) links a point $\mathbf{x} \in \min_{\Psi_{L,R}}(C)$ to a point $\mathbf{y} \in \max_{\Psi_{L,R}}(C)$ into the same lobe. Otherwise, for $C(U, K) > 0$ Lorenz equations link a minimum $\mathbf{x} \in \min_{\Psi_{L,R}}(C)$ of $\Psi_{L,R}$ into a maximum $\mathbf{y} \in \max_{\Psi_{R,L}}(C)$ of the opposite lobe.

Outside $\Xi$, (more precisely into the region $\overline{(\Psi_L \cup \Psi_R) \cap \Xi}$), $2L > G$, $W < 0$ and Casimir sphere continually implodes; here dissipation constrains motion to be globally bounded inside a sphere of radius $R_{max} = \rho + \sigma$, being $Tr(-\Lambda) < 0$.

For $C(U,K)<0$, dynamics links a point $y \in \max_{\Psi_{L,R}}(C)$ to a point $x \in \min_{\Psi_{L,R}}(C)$ in the same lobe; for $C(U,K)>0$, instead, a point $y \in \max_{\Psi_{L,R}}(C)$ will be linked to a point $x \in \min_{\Psi_{R,L}}(C)$ of the opposite lobe. **Fig.10** shows these links.

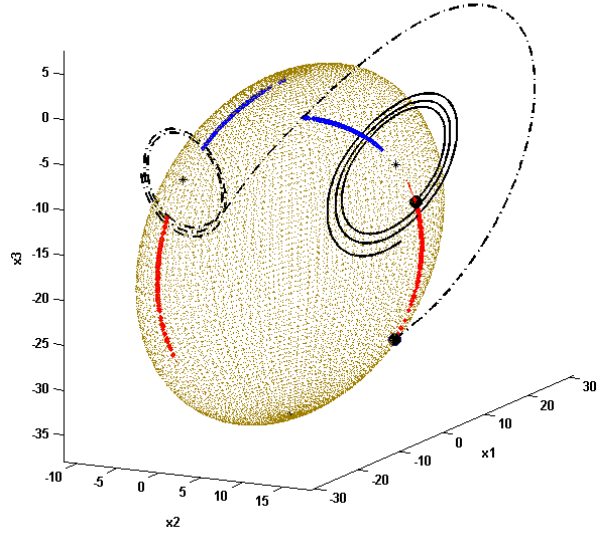

**Fig. 10** Traps and jumps. Trajectories start from right lobe $\Psi_R$ Casimir maxima (shown in red). Solid line track ends up on the same set while dash-dot one reaches maxima on the other lobe $\Psi_L$. Black spots indicate initial conditions for both the trajectories, minima for Casimir are shown in blue, while black stars represent fixed points $\mathbf{x}_\pm, \mathbf{x}_0$

It is remarkable to note that using formalism above described, it is possible to proof the outward spiral motion around fixed points $\mathbf{x}_\pm$.

For a trajectory $l(\mathbf{p}_1, \mathbf{p}_2) \subset \Psi_{L,R}$ linking two points $\mathbf{p}_1(t_1), \mathbf{p}_2(t_2) \in \max_{\Psi_{L,R}}(C)$

$$\int_{t_1}^{t_2} C(U,K)d\tau < 0 \Rightarrow \int_{t_1}^{t_2} (\dot{U} + \beta U - f\omega)d\tau < 0 \quad (31)$$

After integration and indicating by $\omega \bar{x}_3$ the time average of potential energy along $l(\mathbf{p}_1, \mathbf{p}_2)$ we get

$$\frac{x_3(\mathbf{p}_2) - x_3(\mathbf{p}_1)}{t_2 - t_1} < \beta\left(-(\rho+\sigma) - \bar{x}_3\right) < 0 \quad (32)$$

and then $x_3(\mathbf{x}_\pm) > x_3(\mathbf{p}_1) > x_3(\mathbf{p}_2) > x_3(\mathbf{x}_0)$.

In order to better understand the statistics of persistence in the lobes, let's consider the effect of conversion factor (12). System (6) can be written as a particular case of the following set of equations

$$\begin{cases} \dot{x}_1 = -\sigma x_1 + \omega x_2 \\ \dot{x}_2 = -x_1 x_3 - \omega x_1 - x_2 \\ \dot{x}_3 = x_1 x_2 - \beta x_3 - \beta(\rho + \omega) \end{cases} \quad (33).$$

**Fig.11** shows that under variation of parameter $\omega$ the resulting attractor, while conserving its global topology and fractal properties, will explore a greater volume domain due to the increased external forcing.

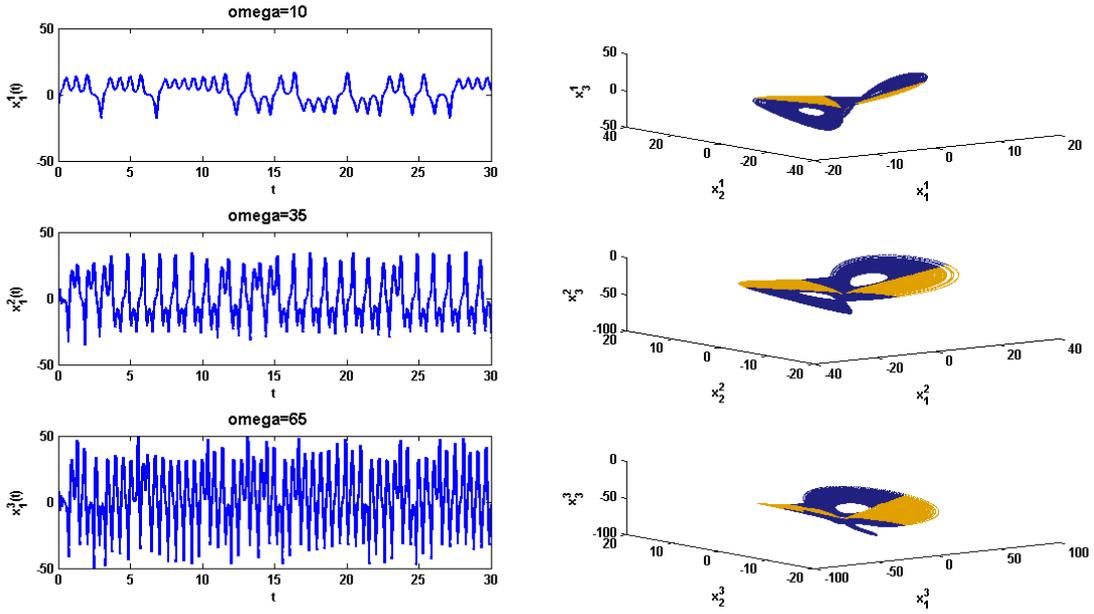

**Fig.11** Statistical behaviour as a function of $\omega$ term. Left: from top to bottom (as $\omega$ increases), $x_1^k(t), k=1,2,3$ shows less and less persistence of trajectory inside a lobe; Right: signs of conversion term over the attractor: regions where $\boldsymbol{C}(U,K) > 0$ (in orange) approach unstable points $\mathbf{x}_\pm$ as $\omega$ increases and region where $\boldsymbol{C}(U,K) < 0$ (in blue) decreases in size.

The most important effect, however, is given by a significant change in persistence statistics; more precisely regions $\boldsymbol{C}(U,K) > 0$ will expand to inner regions of attractor close to fixed points $\mathbf{x}_\pm$, increasing the probability of jumping to the opposite lobe.

## VII. CONCLUSIONS

Up to now, Lorenz system has been studied under many viewpoints in literature; in this paper energy cycle approach has been fully exploited. The nature of Lie-Poisson structure in Lorenz equation has been shown to be fruitful, for example in finding a geometrical invariant, ellipsoid $\Xi_0$, whose physical meaning is the boundary of action between forcing and dissipation. In this manner, kinetic-potential energy transfer term $\boldsymbol{C}(U,K)$ keeps track of dynamical behaviour of trapping and jumping, also giving information about global predictability of the system as illustrated by direct comparison of conversion factors with classical results on predictability.

**Acknowledgements.** Authors warmly thank Prof. E.Kalnay for having provided codes in ref.[20] in order to compute breeding vector analysis.

**REFENCES**

[1]  E.Lorenz, Tellus **7**, 157-167 (1955)
[2]  V.I. Arnold, Proc.Roy. Soc., A **434** 19-22 (1991)
[3]  A.d'Anjous, C.Sarasola and F.J. Torrealdea, Journ. of Phys.:Conference Series 23 238-251. (2005)
[4]  J.E. Marsden and T. Ratiu, *Introduction to Mechanics and Symmetry*, Springer, Berlin, 1994
[5]  P.J. Morrison, Rev. Mod. Phys. **70** 467-521 (1998)
[6]  A.Pasini, V.Pelino and  S.Potestà, Phys.Lett. A **241** (1998) 77-83
[7]  V.Zeitlin, Phys.Rev.Lett **93** No 26 264501-1-264501-3 (2004)
[8]  V.Pelino and A.Pasini, , Phys.Lett. A **291** 389-396 (2001)
[9]  V.Pelino and F.Maimone, Phys Rev.E **76,** (2007)
[10]  V.I. Arnold and B.A. Khesin, *Topological methods in Hydrodynamics*, Springer, Berlin 1988
[11]  A.Pasini and V.Pelino, Phys.Lett. A **275**  435-446 (2000)
[12]  E.N. Lorenz, J.Atmos. Sci. **20** 130 (1963)
[13]   A.Elipe,V. Lanchares, Cel. Mech Dyn Astr **101** 49-68 (2008)
[14] M.Gianfelice,F.Maimone, V.Pelino, S.Vaienti, *Invariant densities for expanding Lorenz-like maps,*  in preparation.
[15]  K.Suzuki,Y.Watanabe,T.Kambe, J.Phys.A Math.Gen **31** 6073-6080 (1998)
[16]  A.Elipe, M.Arribas and A.Riaguas, J.Phys.A:Math.Gen. **30** 587-601 (1997)
[17] R.Salmon, *Lectures on Geophysical Fluid Dynamics* Oxford (1998)
[18]  P.J. Morrison, Journal of Physics **169** 1-12 (2009)
[19] J.Marshall, R.A. Plumb, *Atmosphere, Ocean and Cliamte Dynamics*, Academic Press (2007)
[20]  E.Evans, N.Batti,J.Kinney,L.Pann,M.Pena,S.Chih,E.Kalny,J.Hansen, BAMS 519-524 (2004)
[21] L.A.Smith, C.Ziehmann, K.Fraedrich Q.J.R Meteorol.Soc. **125**, 2855-2886 (1999)